\documentclass[onecolumn,usenatbib]{mnras}
\usepackage{graphicx}
\newcommand{\Dpar}{D_{\parallel}}
\newcommand{\Dperp}{D_{\perp}}
\newcommand{\te}{\'e}
\newcommand{\lskip}{\vskip \baselineskip}
\newcommand{\nskip}{\lskip \noindent}
\newcommand{\bm}[1]{\mbox{\boldmath$ #1 $}}
\newcommand{\grad}{\mbox {\boldmath $\nabla$}}
\newcommand{\bdot}{\mbox{$\bm{\: \cdot \:}$}}
\newcommand{\halfskip}{\vskip 0.5\baselineskip}
\newcommand{\be}{\halfskip \begin{equation}}
\newcommand{\ee}{\end{equation} \halfskip \noindent}
\newcommand{\ba}{\halfskip \begin{eqnarray}}
\newcommand{\ea}{\end{eqnarray} \halfskip \noindent}

\newcommand{\hatr}{\bm{\hat{r}}}
\newcommand{\hatn}{\bm{\hat{n}}}
\newcommand{\hatphi}{\bm{\hat{\phi}}}
\newcommand{\hatb}{\bm{\hat{b}}}
\newcommand{\xt}{(\bm{x} \: , \: t)}

\begin{document}
\title[Anisotropic propagation of Galactic cosmic rays]
{Propagation of Galactic cosmic rays: the influence of anisotropic diffusion}
\author[AL-Zetoun \& Achterberg]{A. AL-Zetoun$^1$\thanks{E-mail: a.al-zetoun@astro.ru.nl}  \&     A. Achterberg$^1$\thanks{E-mail: a.achterberg@astro.ru.nl} \\  
$^1$Department of Astrophysics, IMAPP, Radboud University, Nijmegen, P.O. Box 9010, 6500 GL Nijmegen, The Netherlands}

\date{Accepted..., Received...; in original form ...}
\pubyear{2018}

\maketitle
\begin{abstract}
{ 

We consider the anisotropic diffusion of cosmic rays in the large-scale  Galactic magnetic field, 
where diffusion along the field and diffusion across the field proceeds at different rates.  
To calculate this diffusion we use stochastic differential equations to describe the cosmic ray propagation, solving these numerically. 
The Galactic magnetic field is described using the Jansson-Farrar model for the Galactic magnetic field. 
In this paper we study the influence of perpendicular diffusion on the residence time of cosmic rays in the Galaxy. This provides 
an estimate for the influence of anisotropic diffusion on the residence time and 
the amount of matter (grammage) that a typical cosmic ray traverses during its residence in the Galaxy. 
}
\end{abstract}

\begin{keywords}
Methods: numerical -- diffusion -- magnetic fields -- cosmic rays
\end{keywords}

\section{Introduction}
\label{intro:1}

The theory of cosmic ray (CR) propagation in the Galaxy is a fundamental topic in high-energy astrophysics. 
The long CR residence time in the Galaxy, about \textcolor{black}{$10^{8}$} years at GeV energies (e.g. \citet{StrongM2010}), shows that
CRs propagate diffusively inside the Galaxy and the extended CR halo around the Galactic disk. The precise mode of diffusion is important 
in order to model the propagation, see for instance \citet{B.Potgieter2008}. \textcolor{black}{It influences (among others) the total production of secondary
nuclei by spallation on the intergalactic gas that determines the chemical composition of the CR population.}
Generally, one expects that the presence of the large-scale (ordered) Galactic magnetic field (GMF) leads to 
anisotropic diffusion since its magnitude is comparable to the smaller-scale random field component.
In that case diffusion along the large-scale field
(with diffusion coefficient $\Dpar$) and diffusion in the plane perpendicular to the field (with diffusion coefficient $\Dperp$) proceeds at different rates.
Quite generally one expects $\Dperp \le \Dpar$. 
\nskip
Parallel diffusion is usually described using the quasilinear theory (e.g. \citet{Wentzel74}) of wave-particle interactions, 
where scattering by Alfv{\te}n waves or -more generally- MHD turbulence 
leads to a finite CR mean-free path along the magnetic field. 
Perpendicular diffusion is more difficult to describe: it can also be due to scattering by waves, (e.g. \citet{Schlick2002}, Ch. 12). 
But in addition there is the effect of the stochastic wandering individual magnetic field lines with respect to the mean field (e.g. \citet{Jokipii66}).
The effective perpendicular diffusion coefficient $D_{\perp}$ then is a combination of both processes. 
This complicates the theory. Recently, perpendicular diffusion has been described using the more complicated 
nonlinear guiding center theory, see for instance \citet{Shalchi2009}, \citet{ShalchiB2010},
and \citet{GaSha2017}. 
These authors derive a ratio between the perpendicular and parallel diffusion coefficients of order $\Dperp/ \Dpar = 10^{-4}- 10^{-1}$.  
We will assume in what follows
that diffusion is isotropic in the plane perpendicular to the large-scale field so that the diffusion tensor has the simple form 
$\bm{\mathrm{D}} = {\rm diag}(\Dperp \: , \: \Dperp \: , \: \Dpar)$ in a local coordinate frame with base vectors 
$\hat{\bm{e}}_{1} \: , \: \hat{\bm{e}}_{2} \: , \: \hat{\bm{e}}_{3}$ with
$\hat{\bm{e}}_{3}$ aligned with the direction of the large-scale magnetic field. 
We will also assume that the
turbulence responsible for the diffusion is axi-symmetric around the field direction so that there are 
no off-diagonal (mixed) components in $\bm{\mathrm{D}}$. Other choices with fully anisotropic diffusion are possible, see for instance \citet{Effenb}. 
\nskip
Typical values for the magnitude of the diffusion coefficients inside the Galaxy are usually derived by fitting CR
composition data in an axi-symmetric model,  assuming (by default) radial diffusion. 
One finds $D\simeq (3-5) \times 10^{28}$ cm$^{2}$ s$^{-1}$ at a CR energy $\simeq 1$ GeV per nucleon, see for instance \citet{StrongM2007}. 
This agrees with the typical residence time $t_{\rm res} \simeq 10^{8}$ years that is calculated from measured isotope ratios for radioactively unstable/stable isotopes 
of Beryllium and Aluminum. 
\nskip
If one assumes that CRs diffuse perpendicular to the disc of the Galaxy, escaping once they reach the upper (lower) edge of the cosmic ray disk at height $ z = \pm H_{\rm cr}$ 
above the Galactic plane. \textcolor{black}{The estimated value of $ H_{\rm cr}$ in different CR propagation models,  
such as GALPROP, see \citet{StrongM2010}, DRAGON, see \citet{Evoli2008} and CRPropa, see \citet{Batista}. It generally falls in the range $H_{\rm cr} \simeq 4-10$ kpc. 
In this work, and as a first test of the method, we have chosen a value of $H_{\rm cr} \simeq 1$ kpc.  One has a typical CR residence time in the Galaxy equal to
$t_{\rm res}\simeq H_{\rm cr}^{2} /2 D$, where $D$ is the diffusion coefficient. } Using typical values \textcolor{black}{for CRs with an energy of $\sim 1$ GeV per nucleon} one finds
\be
	D \simeq \frac{H_{\rm cr}^2}{2 t_{\rm res}} =  1.5 \times 10^{27} \: \left( \frac{H_{\rm cr}}{{\rm 1 \;\rm kpc}} \right)^2 
	\: \left( \frac{t_{\rm res}}{\textcolor{black}{10^{8}} \; {\rm years}} \right) \; {\rm cm}^2 \: {\rm s}^{-1} \; .
\ee 
The aforementioned measurements also
indicate that the residence time varies with rigidity $E/Z$, with $Z$ the charge number,  as $t_{\rm res} \propto (E/Z)^{s}$ with $s \simeq 0.4-0.6$.
\nskip
There are a number of codes that model Galactic CR propagation and the CR interactions (losses, spallation, \textcolor{black}{decay of unstable nuclei and} re-acceleration)
in the Galaxy, such as GALPROP, see \citet{StrongM2010}, DRAGON, see \citet{Evoli2008} and CRPropa, see \citet{Batista}. 
\textcolor{black}{In this paper we do not take into account the effects of CR advection by the Galactic winds, energy losses, and re-acceleration. We also do not
consider the network of nuclear reactions that leads to the production of secondary CRs.
These effects will be considered in future work}. 
\nskip
However, \textcolor{black}{in these codes one usually limits}
the calculations to models with cylindrical symmetry models with an effective diffusion coefficient $D$, 
thereby neglecting the influence of the geometry and strength of the GMF on CR diffusion. 
Recently some effort has been taken to consider anisotropic diffusion models for Galactic cosmic rays, see \citet{Evolietal}.
\nskip
We will consider anisotropic diffusion of CRs inside the Galaxy.
We note that a similar calculation, again  employing stochastic differential equations, has been  applied to the interplanetary transport and 
acceleration of Solar energetic particles in the Solar wind see \citep{WDroge2010} and  
the references therein.
\nskip
The outline of this paper is as follows: in Section \ref{sec:2} we discuss the Galactic magnetic field and introduce the main features of the Jansson-Farrar. 
In Section \ref{sec:3} we present our numerical solution method to the  CR diffusion, based on the It\^o formulation of stochastic differential equations, 
using relevant input, like the diffusion tensor and its connection to the galactic magnetic field in Section ${\color{blue} \;3.1}$. 
The grammage calculation is discussed in Section \ref{sec:3.2}. 
In Section \ref{sec:4} we present the spatial distribution of the CRs from a single source the associated residence time distribution, as well as the grammage of CR particles.
We do this for various combinations of parallel and perpendicular diffusion. 
Section \ref{sec:5} discusses the implications of the results presented in this paper and summarizes our conclusions.
\section{Model for the Galactic magnetic field}
\label{sec:2}

The magnetic field in the diffuse interstellar medium (ISM) has two distinct components: a large-scale ($ \rm kpc$) regular field, as well as a small-scale turbulent component. 
The strength of the total magnetic field in the Galaxy has been estimated of about 6$ \pm 2 $  $\mu$G, see \citet{Beck2001}, and \citet{Haverkorn2015}, with the
amplitude of the random field comparable to that of the large-scale field.  The large-scale GMF affects the propagation of CRs  \citet{StrongM2007},
the dynamics of molecular clouds and star formation, see \citet{HennebelleF2012}. 
\nskip
In this paper we  use the GMF model of \citet{JF2012A}, and \citet{JF2012B} for this
large-scale (regular) field. This model 
matches the WMAP7 22 GHz synchrotron intensity map and the Galactic Faraday rotation data obtained from polarization measurements of background sources.
\nskip
The precise structure of the random component of the GMF on the wide range of scales relevant for CR propagation 
(from sub-pc scales to kpc scales) is not very well known. It depends on the details of the Galactic dynamo responsible for generating and maintaining these fields, see for instance
\citet{RuzmaikinS1988}, \citet{Kulsrud1999}, and \citet{Shukurov2004}. There is no consensus yet on the best description for the dynamo process.
Given that uncertainty we will not attempt here to model in detail the processes leading to diffusion along and across
the large-scale field. We will limit ourselves to a parameter study, focusing on the effect of using different values 
for the ratio of the perpendicular and parallel diffusion coefficients, $\Dperp/\Dpar$.
\nskip 
\subsection{The Jansson-Farrar model}
\label{sec:2.1}

We employ the GMF model of  \citet{JF2012A}, and \citet{JF2012B} to describe the large-scale Galactic field. We briefly list its main features here.
The model uses Galacto-centric $(r\:,\: \phi\:,\: z)$ cylindrical coordinates as well as a right-handed Cartesian $(x\:,\: y\;,\; z)$  coordinate system. The Sun is located
along the negative $x$-axis, at $x=-8.5\; \rm kpc$. The galactic north is in the positive $z$-axis. 
The GMF is set to zero for $r > 20\; \rm kpc$, and within a $1\; \rm kpc$ radius sphere centered on the Galactic center. The model ensures explicitly  the vanishing divergence of the magnetic field: $\grad \bdot \bm{B} = 0$.  We summarize the main properties of the Jansson-Farrar model \textcolor{black}{in} what follows.
\nskip
The Jansson-Farrar model has three independent components: 
\begin{enumerate}
\item
The {\em disk field} extends for a Galacto-centric radius $ 3\: {\rm kpc} < r\:< 20\: {\rm kpc}$ and is parallel to the Galactic plane. It is uniform
in height, extending from $z = 0$ to $z = \pm H_{\rm disk} = 0.4\; \rm kpc$.
In the inner molecular ring $(3\; {\rm kpc} \:<\:r\:<\:5\:{\rm kpc})$ the field is purely azimuthal with a field strength $B_{\rm ring}$ = 0.1 $\mu$G. 
For $ 5\; {\rm kpc}\;<\;r\;<\;20\;{\rm kpc}$  the disk field is composed of different sections of a logarithmic spirals 
(eight in total) each with a different field strength and with a constant pitch angle. 
The magnetic field in the spiral sections is defined by
\be
	\bm{B}_{\rm disk}(r \: , \: \phi) = B_{r}(r) \: \hatr + B_{\phi}(r) \:\hatphi \; . 
\ee 
The pitch $p$ of the disk field is
\be
\label{pitch}
	p = \tan^{-1} \left( \frac {B_{r}}{B_{\phi}} \right) = 11.5 \; {\rm degrees} \; .
\ee
Magnetic flux conservation demands that $r \: \Delta \phi \: B_{r} =$ constant, where $\Delta \phi$ is the constant range in azimuth of a section of the 
disk field section. 
This implies that the disk field decays with Galacto-centric radius $r$ as
\be
	| \bm{B}_{\rm disk} | \propto r^{-1} \; .
\ee
The typical strength of this field is a $\sim 5 \mu {\rm G}$ at $r = 5$ kpc, but varies considerably in the different sections, see Table 1 of \citet{JF2012A}.
The vertical extent of the disk field above (and below) the stellar disk is governed by a transition of the form 
$B_{\rm disk}(z) = B_{\rm disk}(0) \times \left[1 - L(z \: , \: h \: , \: w) \right]$ where
\be
\label{transcomp1}
	L(z \: , \: h \: , \: w) \equiv \left(1 + e^{2(|z|-h)/w} \right)^{-1} \; ,
\ee
with $h = 0.4$ kpc and $w = 0.27$ kpc.
\nskip
\item
The {\em halo field} is a purely toroidal component with a different radial and vertical extension in the northern and southern halo. 
The halo field is
$\bm{B}_{\rm halo} = B(r\:,\:z) \: \hatphi$, with
\be
	B(r\:,\:z) = B_{0{\rm h}} \:L\:(z\:,\: h \;,\: w)\;[\:1-L\:(r\:,\:r_{\rm h}\:,\:w_{\rm h}\:)\:] \: {\rm exp}(-\frac{z}{z_0}) \; .
\ee
This field applies above the disk-halo transition height $ \left|z  \right|=H_{\rm disk}  \simeq   0.4\; \rm kpc$. 
The vertical scale height of this field is $z_0 \simeq 5.3\; \rm kpc$. The value of $B_{0 \rm h}$ is different for the southern and northern hemisphere
\be
	B_{0 \rm h} = \left\{
			\begin{array}{rcl}
          +1.4\: \mu {\rm G} & ; & \mbox{Northern halo} \; , \\
          & & \\
          -1.1\: \mu {\rm G} & ; & \mbox{Southern halo} \; .
			\end{array}
			\right. 
\ee
The function $L(r\:,\:r_{\rm h}\:,\: w_{\rm h}) $ is defined as (compare Eqn. \ref{transcomp1})
\be
	L(r\:,\:r_{\rm h}\:,\: w_{\rm h}) = \left[ 1 + {\rm exp}\left(-\frac{2(r-r_{\rm h})}{w_{\rm h}}\right) \right]^{-1} \; ,
\ee
with a width $w_{\rm h}\simeq 0.2\; \rm kpc$, a transition radius $r_{\rm h} \simeq 9.22\; \rm kpc$ in the northern halo, 
and $r_{\rm h} > 16\; \rm kpc$ in the southern halo.

\nskip
\item
The magnetic field component perpendicular to the plane of the Galactic disk is the so-called {\em X-field}. 
This component is  azimuthally symmetric, and the shape of the field lines is specified at any position $(r\:,\: z) $  
in terms of  the cylindrical radius $r_{\rm p}$ at which the field line crosses the Galactic plane $z$ = 0. 
Individual field lines in the X-field are straight, with components that can be written as 
\be 
	B^{X}_{r} = B^{X} \: \cos(i) \;\;\;,  \;\;\;\;      B^{X}_{z} = B^{X}\: \sin(i) \; .
\ee 
The elevation angle $i$ varies with radius $r_{\rm p}$ as 
\be
\label{tan:inclination}
	\tan(i) =  \left\{\begin{array}{ccl}
              \tan(i_{0}) \: (\: r_{X}/r_{\rm p}\:) & ; &r_{\rm p} < r_{X}= 4.8\: {\rm kpc} \; , \\
               & & \nonumber\\
               \tan(i_{0}) & ; & r_{\rm p} \geq r_{X}= 4.8\: {\rm kpc} \; . \\ 
	\end{array} \right.
\ee
The elevation angle $i_{0}$ that appears here equals $i_{0} = 49^{\rm o}$. 
In the Galactic mid-plane the strength of the X-field is assumed to vary as
\be
	B^{X} (r_{\rm p}\:,\: z = 0) = B_{0}\: {\rm exp}(-r_{\rm p}/ H_{X}) \; , 
\ee 
where $H_{X} = 2.9\; \rm kpc$, and $B_{0} = 4.6\:\mu$G. Flux conservation implies
\be
	2\pi\: r_{\rm p}\:{\rm d}r_{\rm p} \; B^{X}_{z}(r = r_{\rm p} \: , \: z = 0) = 2\pi\: r \:{\rm d}r \: B^{X}_{z}(r) \; .
\ee
The radius $r(z)$ along a given field line projected onto the $r-z$ plane follows from 
\be
	\frac{{\rm d} z}{{\rm d} r}= \frac{B_{z}^{X}}{B_{r}^{X}} = \tan(i) \; ,
\ee
where the elevation angle $i$ is constant on each individual field line. 
The solution for $r$ along a given field line that satisfies $ r = r_{\rm p}$ at $z = 0$ together 
with equation (\ref{tan:inclination}) is
\be
	r - r_{\rm p} = \frac {z}{\tan(i)} = 
            \left\{
             \begin{array}{rcl}
                  \frac {\displaystyle r_{\rm p}\left|z \right|}{\displaystyle r_{X}\: \tan(i_{0})} & ; &r_{\rm p} < r_{X} \; , \\ 
                  & & \nonumber\\
                 \frac{\displaystyle \left|z \right|}{\displaystyle \tan(i_{0})} \;\; & ; &r_{\rm p} \geq r_{X} \; . \\
             \end{array}
              \right.  
\ee
This determines $r(r_{\rm p}\:, \: z)$ on a given (projected) field line. From these relations we get
\be
	\frac{{\rm d}r}{{\rm d}r_{\rm p}} = 
           \left\{
                  \begin{array}{rcl}
                      {\displaystyle 1+ \frac{ \left|z \right|}{r_{X}\: \tan(i_{0})} = \frac{r}{r_{\rm p}}} & ; &r_{\rm p} < r_{X} \; , \\
                      & & \nonumber\\
                     1\;\;\;\;\;\;\;\;\;\;\;\;\;\;\;\;\;\ & ; &r_{\rm p} \geq r_{X} \; . \\
                   \end{array}
                    \right. \;  
\ee
The flux conservation condition (14) becomes
\be
\label{Xfield}
	B_{z}^{X} (r\:,\:z)= B_{z}^{X}(r_{\rm p})\; \bigg[\frac{r\: {\rm d}r}{r_{\rm p}\: {\rm d}r_{\rm p}}\bigg]^{-1} =  B_{z}^{X}(r_{\rm p}) \:  
	\left\{
	\begin{array}{rcl}
               {\displaystyle \left( \frac{r}{r_{\rm p}} \right)^{-2}} & ; &r_{\rm p} < r_{X} \; , \\
                & & \nonumber\\
                {\displaystyle \left( \frac{r}{r_{\rm p}} \right)^{-1}} & ; &r_{\rm p} \geq r_{X} \; . \\
	\end{array}
	\right.\;
\ee
Since $\tan(i)$ is constant on each filed line (but varies from field line to field line for $r_{\rm p} \: < r_{X}$) the same scaling holds for $ |\bm{B}^{X}| (r\:,\:z)$. 
For more details about the model  see \citet{JF2012A} and  references therein.
\end{enumerate}

\section{Cosmic ray diffusion: the numerical method}
\label{sec:3}

To describe diffusion we employ a numerical method based on the It\^o formulation of stochastic differential equations, see \citet{Oks1992}. This method can be applied
to diffusion processes in general, see for instance \citet{KruAcht94}. \textcolor{black}{This method has become popular in CR propagation. The main benefit of this method it is straightforward parallelization and scalability, and that it can tackle the very complicated problems that are usually associated by solving differential equations of this type.} 
\nskip
\textcolor{black}{Recently, the PICARD code  has been developed \citet{kissmann}.  
PICARD uses advanced contemporary numerical algorithms for solving the diffusion equation in order to efficiently compute the {\em steady state} solution
for the equations of CR  propagation. 
Our method -by design- employs time-stepping and therefore inherently considers the time-dependent case. Steady solutions are (in principle) obtained after a large
number of time steps, assuming of course that none of the model parameters (injection, field geometry etc.) changes in time.} 
\nskip
The example of a one-dimensional random walk with diffusion coefficient
$D$ \textcolor{black}{may serve a simple explanation for the method that we employ here.} 
Consider the following prescription that advances a particle along the $x$-axis over a distance $\Delta x$
in a timespan $\Delta t$, from $t$ to $t + \Delta t$ 
(e.g. \citet{Higham2001})
\be
\label{Itostep}
 	\Delta x = V(x \: , \: t) \;\Delta t +\sqrt {2D(x \: , \: t)} \: W(\Delta t) \; .
\ee
Here $V(x \: , \: t)$ is a mean velocity (see below \textcolor{black}{for a precise definition}) and $W(\Delta t)$ is a (random) Wiener process, 
drawn from the following Gaussian probability distribution
\be
\label{probdistr}
	P(W \: | \: \Delta t) = \frac{1}{\sqrt{2\pi \: \Delta t}} \: {\rm exp}(-W^{2}/2 \: \Delta t)\; .
\ee
With probability distribution function (\ref{probdistr}) the average step size $< \Delta x >$ and its dispersion satisfy
\be
	\left< \Delta x \right> = V \: \Delta t \; \; , \; \; \left< (\Delta x)^2 \right> - \left< \Delta x \right>^2 = 2 D \: \Delta t \; .
\ee
This obviously represents a drift at velocity $V$ together with diffusion. The drift velocity $V$ should be defined as
\be
\label{driftvelo}
	V(x \: , \: t) = U(x \: , \: t) + \frac{\partial D(x \: , \: t)}{\partial x} \; .
\ee
Here $U$ is the mean velocity of the medium and the $\partial D/\partial x$ gradient term corrects for the lopsidedness of diffusion that occurs 
when the diffusion coefficient depends on position.
One can show that simulating a large number of particles with this prescription leads to a distribution $f(x \: , \: t)$ of particles along the $x$-axis,
so that $f(x \: , \: t) \: {\rm d}x$ is the probability of finding a particle in the interval $x$, $x + {\rm d}x$ at time $t$,
that satisfies an advection-diffusion equation of the \textcolor{black}{conservative form as used in} \citet{AchtKru92}
\be
\label{diffeqn}
	\frac{\partial f(x \: , \: t)}{\partial t} + \frac{\partial}{\partial x} \left[ \: U(x \: , \: t) \: f(x \: , \: t) \: \right] = 
	\frac{\partial}{\partial x} \left( D(x \: , \: t) \: \frac{\partial f(x \: , \: t)}{\partial x} \right) \; .
\ee
This method is straightforwardly generalized to more dimensions and to anisotropic diffusion, and is also 
easy to implement numerically.  \textcolor{black}{In what follows we neglect advection by a Galactic wind and the effects of field-line curvature leading
to cross-field drift, putting $U = 0$. They will be included in later calculations. 
For the purpose of practical calculations} 
it is convenient to replace the random step in prescription (\ref{Itostep}) by \textcolor{black}{the prescription}
\be
\label{diffstep}
	\Delta x_{\rm diff} \equiv \sqrt {2D(x \: , \: t)} \: W(\Delta t) = \sqrt{2D(x \: , \: t) \: \Delta t} \: \xi \; ,
\ee
where $\xi$ is a random variable, chosen at each new each time step from a distribution $P(\xi)$ that has zero mean ($< \xi > = 0$) and unit dispersion ($<\xi^2> = 1$). 
This approach is equivalent to defining a new random walk, with the same diffusion coefficient but with a different step size and step frequency.
\textcolor{black}{This also illustrates one of the main strengths of this method: as long as one chooses the typical diffusive step size 
$\sim \sqrt{2D(x \: , \: t) \: \Delta t}$ smaller than the
(macroscopic) scales on which the properties of the interstellar medium, source distribution and magnetic field change (typically on $\sim$ kpc scales), 
one can use a step size much larger
than the actual step size (typically $ 0.1-1$  pc) of the underlying random walk and still obtain the correct CR distribution. This speeds up the calculation
by a large amount. Of course, like all \lq{}particle-based\rq{} methods a sufficient number of particles needs to be used in order to overcome the problem of Poisson noise.}

\subsection{Anisotropic diffusion in a magnetic field}
\label{sec:3.1}

CR diffusion in a large-scale (regular) magnetic field is -generally speaking- anisotropic, e.g. \citet{GiacaloneJ1999}, and references therein. 
Typically CRs diffuse rapidly along the field lines, with slower diffusion across the magnetic field.
As mentioned briefly in the Introduction, we choose the diffusion tensor $\bm{\mathrm{D}}$ to be diagonal, 
with diffusion coefficient $\Dpar$ along the magnetic field and with diffusion characterized by a diffusion coefficient
$\Dperp$ in the plane perpendicular to the field. 
The diffusion tensor  $\bm{\mathrm{D}}$ takes the form (in dyadic notation) 
\be
\label{difftensor}
	\bm{\mathrm{D}} = D_{\perp} \: \left( \bm{\mathrm{I}} - \hatb \hatb \right) + \Dpar \: \hatb \hatb \; ,
\ee
with $\bm{\mathrm{I}}$ the $3 \times 3$ unit tensor, and $\hatb \equiv \bm{B}/|  \bm{B} |$ the unit vector along the magnetic field.
The generalization of recipe (\ref{diffstep}) to this particular case is particularly simple: the diffusive step can be represented as
\be
\label{diffstepsize}
	\Delta \bm{x}_{\rm diff} =    \sqrt{2 \Dperp \: \Delta t}\; \xi_{1}\: \hat{\bm{e}}_{1}
	+  \sqrt{2 \Dperp \: \Delta t}\: \xi_{2} \;	\hat{\bm{e}}_{2} + \sqrt{2 \Dpar \:  \Delta t}\; \xi_{3} \: \hatb \; ,
\ee
all in the local Cartesian coordinate system with base vectors 
$\hat{\bm{e}}_{1} \: , \: \hat{\bm{e}}_{2} \: , \: \hat{\bm{e}}_{3}$ with the local large-scale 
magnetic field along $\hat{\bm{e}}_{3}$.
In recipe (\ref{diffstepsize}) the quantities $\xi_{1}$, $\xi_{2}$ and $\xi_{3}$ are three independent (that is: statistically {\em un}correlated) 
stochastic variables, with zero mean and unit dispersion.
\nskip
Diffusion tensor (\ref{difftensor}) neglects the effect of the off-diagonal components in the plane perpendicular to $\bm{B}$. In the simplest collisional
theories for diffusion  (e.g. \citet{MelroseII}, Ch. 7) these components take the form $D_{xy} = - D_{yx} \equiv D_{\rm a}$. 
In that case they lead to an additional drift velocity, but not to true diffusion. 
A similar approach using stochastic differential equations to describe cosmic-ray diffusion
has been taken recently by \citet{Mertenetal} in their low-energy extension of the CRPropa code. 

\subsection{Momentum diffusion and expansion losses in a wind}

For completeness sake we point out that stochastic re-acceleration of CRs during propagation in the Galactic disk can be included
straightforwardly using the same algorithm. Re-acceleration in the interstellar medium is 
usually described by an extra  diffusive term  in particle momentum $p$. In the CR propagation equation the corresponding term takes the form
\be
\label{reaccn}
	\left( \frac{\partial f}{\partial t} \right)_{\rm acc} = \frac{1}{p^2} \frac{\partial}{\partial p} \left( p^2 D_{p} \: \frac{\partial f}{\partial p} \right) \; .
\ee
Here $D_{p}$ is the momentum diffusion coefficient. The acceleration process can be modeled using the same method as employed for spatial diffusion.  As shown in
\citet{AchtKru92}, this is most conveniently done by introducing the dimensionless logarithmic momentum variable $u = \ln(p/mc)$ and using a density
 $n(\bm{x} \: , \: t \: , \: u) \equiv 4 \pi p^3 \: f(\bm{x} \: , \: t \: , \: p) = {\rm d}N/{\rm d}^3 \bm{x} \: {\rm d}u$.
One advances $u$ according to the analogue of relation (\ref{Itostep})
\be
	{\rm d} u = \dot{u} \: {\rm d}t + \sqrt{2 \: D_{u} \: \Delta t} \: \xi_{u} \; ,
\ee
where
\be
	D_{u} \equiv \frac{D_{p}}{p^2} \; \; , \; \; 
	\dot{u} \equiv \frac{\partial D_{u}}{\partial u} + 3 D_{u}  = \frac{1}{p^2} \frac{\partial}{\partial p} \left( p \: D_{p} \right) \; .
\ee
The independent stochastic variable (Wiener process) $\xi_{u}$ again has zero mean and unit dispersion. In the calculations presented in this paper we neglect the effect of re-acceleration, 
effectively putting $D_{u} = 0$. If a Galactic wind is present with velocity $\bm{U} \xt$ the expansion losses due to this wind, $\dot{p} = - (p/3) \: \grad \bdot \bm{U}$, 
can be incorporated by using for $\dot{u}$
\be
	\dot{u} = - \frac{\grad \bdot \bm{U}}{3} + \frac{1}{p^2} \frac{\partial}{\partial p} \left( p \: D_{p} \right) \; .
\ee
Other regular energy losses can be treated in an analogous manner. 

\subsection{Grammage calculation}
\label{sec:3.2}

CRs diffusing through the interstellar gas inside the Galaxy accumulate grammage $\Sigma_{\rm cr} \sim ({\rm pathlength}) \times ({\rm mass \; density})$  
as they (typically) traverse the Galactic disc thousands of times. 
The amount accumulated depends mostly on the mass density $\rho$ of diffuse hydrogen/ helium gas. 
The density of the diffuse gas in the mid-plane of the Galactic disk (i.e. at $z = 0$) scales with radius $r$ as see \citet{Kalb2009}
\be
\label{cdens}
	\rho(r \: , \: 0) = \left\{ \begin{array}{ll}
	\rho_{0} & \mbox{for $r < R_{\rm c} = 7$ kpc}  \; , \\
	& \\
	\rho_{0} \: {\rm exp}(-(r - R_{\rm c})/R_{\rm d}) & \mbox{for $r > R_{\rm c} = 7$ kpc}  \; . \\
	\end{array} \right. 
\ee
The scale length in the exponential is $R_{\rm d} \simeq 3.15$ kpc, and $\rho_{0} \simeq 3 \times 10^{-24} \; {\rm  g} \; {\rm cm}^{-3}$. 
The typical thickness $H_{\rm d}$ of the hydrogen disk flares out as $r$ increases: it scales as
\be
	H_{\rm d}(r) = H_{0} \: {\rm exp}(r/R_{\rm h}) \; ,
\ee
with $R_{\rm h} \simeq 9.8$ kpc and $H_{0} \simeq 0.063$ kpc. It is expected that this law will break down for $r < 5$ kpc, but we will use it anyway for lack of a better model.
In view of this we can adopt an axi-symmetric density distribution that scales as
\be
	\rho(r \: , \: z) = \rho(r \: , \: 0) \: {\rm exp}\left( - \frac{z}{H_{\rm d}(r)} \right) \; .
\ee
As we have  anisotropic diffusion, the grammage seen in a cosmic ray life time depends on the pitch angle $p$ (see prescription (\ref{pitch})) of the magnetic field in the disk. 
The typical amount of matter traversed  by cosmic ray particles during their propagation,  expressed in terms of the grammage  $\Sigma_{cr}$, 
follows from observations of secondary CRs: $\Sigma_{\rm cr} \simeq 15\;{ \rm g\; cm}^{-2}$  $ (\frac{E}{1 \; {\rm GeV}})^{- s}$ with the exponent $s \simeq 0.4-0.6$.
In our calculation the grammage is accumulated each time step by
\be
	\Delta \Sigma_{\rm cr} = \rho(\bm{r}_{\rm cr} ) \: v \: \Delta t \; ,
\ee
with $v = c \: \sqrt{1 - (mc^2/E)^2}$ the velocity of a cosmic ray with rest mass $m$ and energy $E$ and the position of the CR inside the Galaxy 
denoted by $\bm{r}_{\rm cr}$. 
In this paper, and as a first test, we keep $E$ constant, and consider particles with
$E \gg mc^2$ so that $v \simeq c$.
\section{Results of the simulations}
\label{sec:4}

\begin{figure}
\includegraphics [width=\columnwidth]{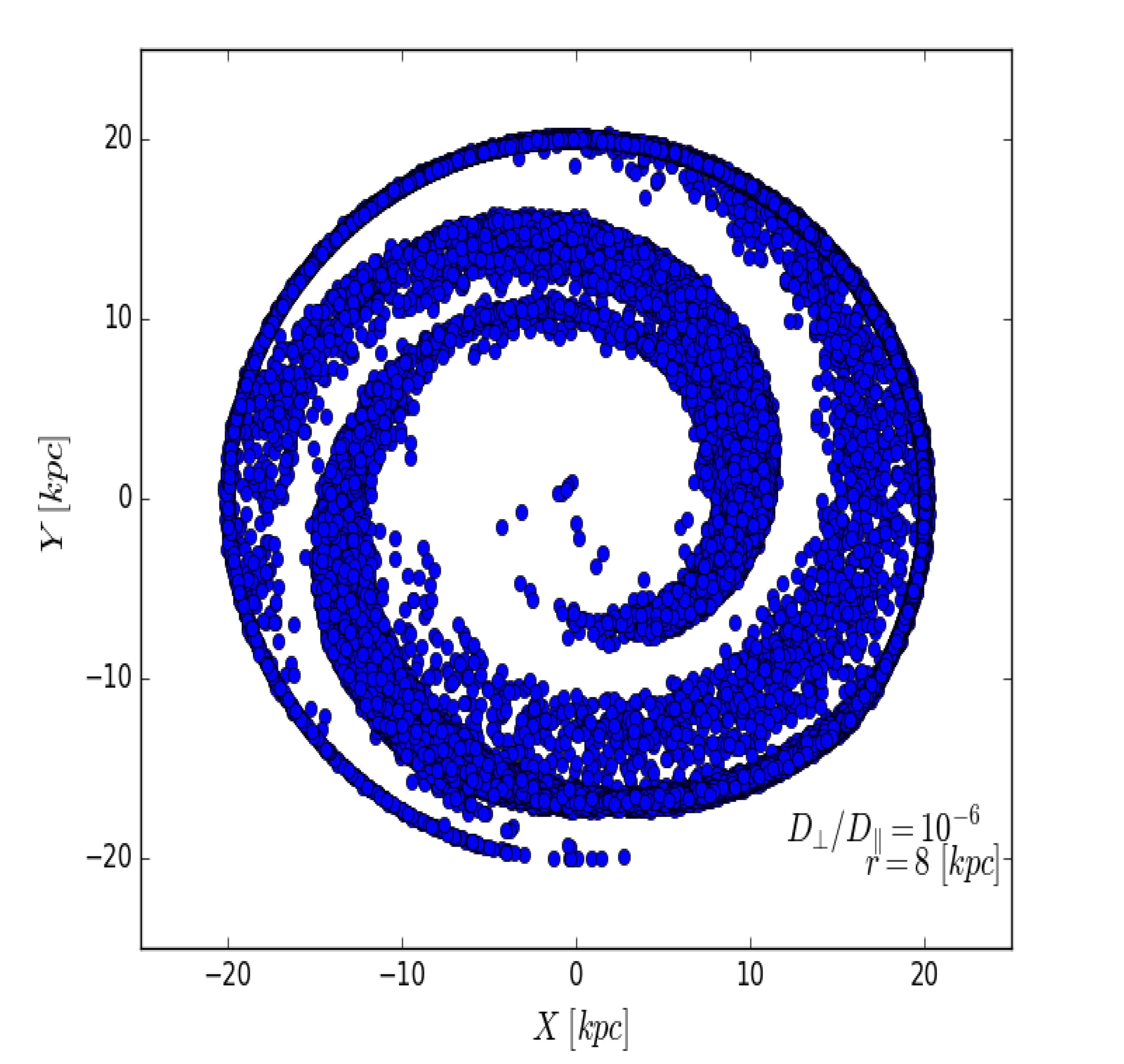}
\caption{This figure shows the position of diffusing CRs in the plane of the Galactic disk for every 1000 steps, 
until they  reach the upper or lower boundary at $ z=\pm H_{\rm cr}$  or $r = r_{\max} = 20$ kpc. 
All particles are injected at the same position, i.e.  at  $X = 8\;\rm  kpc$, $Y = 0$. In this simulation we choose $\Dperp/\Dpar=10^{-6}$, 
an artificially low value in order to check if CRs follow magnetic field lines. 
The parallel diffusion coefficient in the simulation equals $\Dpar = 3 \times 10^{28} \: {\rm cm}^2 \: {\rm s}^{-1}$.
The fact that they do not follow precisely the constant-pitch spiral field lines of the
disk field is due to the inclination of the X-field in the $r-z$ plane in the Jansson-Farrar model. Also, because of the small value of $\Dperp$ the residence time
inside the Galaxy is artificially long compared to more realistic values.}
\label{figure1}
\end{figure}

\begin{figure}
\includegraphics[width=\columnwidth]{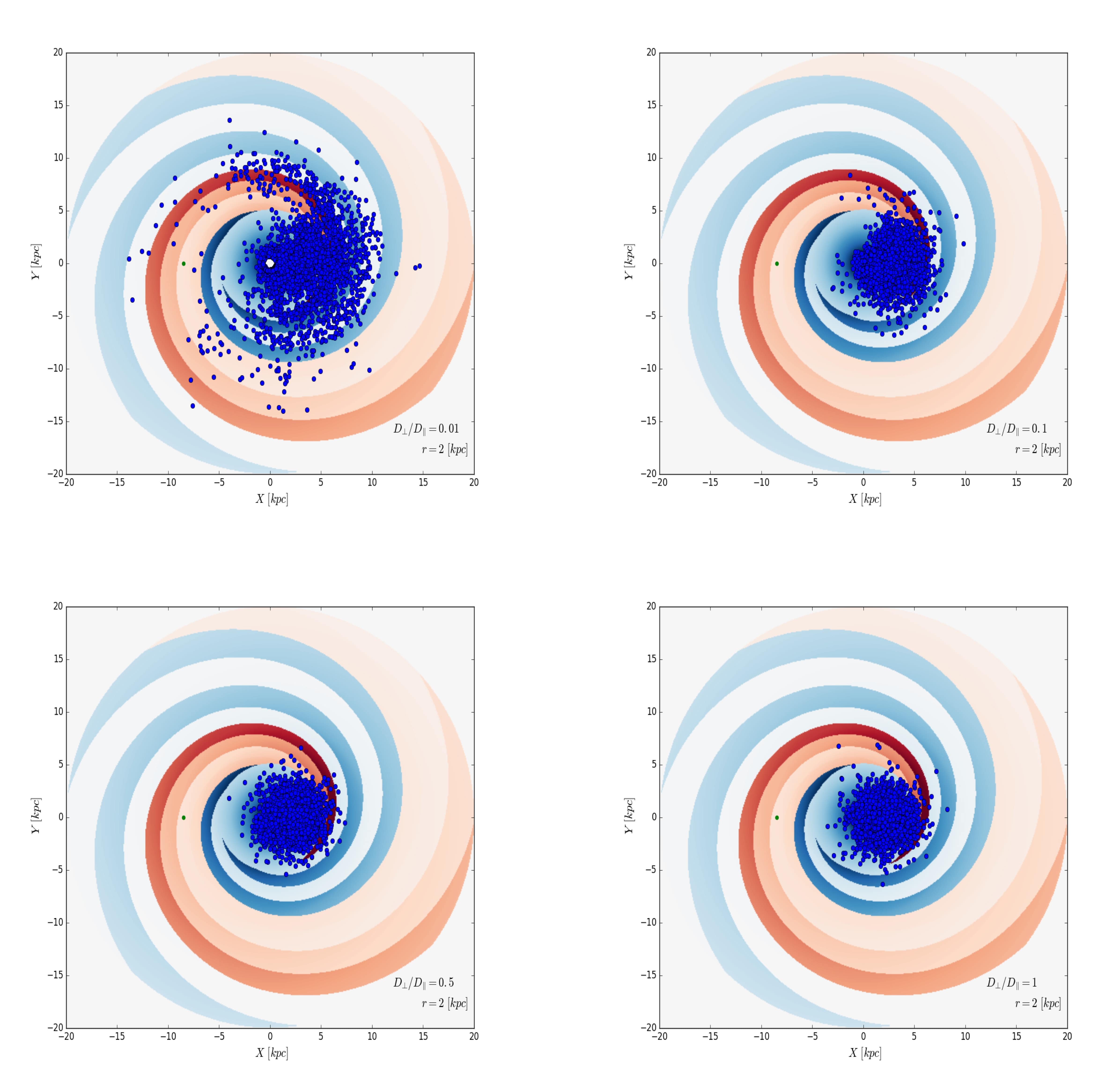}
\caption{This figure shows the position of diffusing cosmic rays in the plane of the disk at the moment they reach the upper or lower boundary at $ z=\pm H_{\rm cr}$. 
All particles are injected at the same position on the positive $x$-axis at  $2\;\rm  kpc$. The four panels show the results for four values of the ratio $ \epsilon =\Dperp/\Dpar $:
$\epsilon = 0.01 \: , \: 0.1 \: , \: 0.5$, and $1.0$. 
The position of the Sun at $(r\;=-8.5\; \rm kpc)$ is marked by the green spot. Color coding is used to indicate the different spiral sections of the Jansson-Farrar Disk Field.
The symmetry of the Jansson-Farrar magnetic field is such that injection on the negative $x$-axis at the same distance from the Galactic center
would not lead to markedly different results. For these values of the parameters escape at the outer radius of the simulation box at $r = 20 \; {\rm kpc}$ plays no
role.} %
\label{figure2}  
\end{figure}

In this Section we describe a first set of simulations with the simplest possible assumptions.
In all cases we inject the particles impulsively (that is: in a burst of negligible duration) in the mid-plane $(z=0)$ of the Galactic disk and 
at a fixed Galacto-centric radius close to that of the Sun $(r\;=R_{\odot}\;=8.5\; \rm kpc)$.  
We then let the CRs diffuse according to prescription (\ref{diffstepsize}), assuming no mean flow. We also assume uniform diffusion coefficients $\Dperp$ and $\Dpar$.
The more realistic case, where $\Dperp$ and $\Dpar$ depend on the local magnetic field strength, the position in the halo and/or particle energy, will be considered in later papers.
\nskip 
The total residence time and accumulated grammage of the CRs are recorded once they reach upper (lower) boundary  of the cosmic ray halo above the Galactic disk, 
located at $z\;=+ H_{\rm cr}$ ($z=- H_{\rm cr} $). 
We take $H_{\rm cr} = 1$ kpc, the typical size of the non-thermal (synchrotron) halo observed in nearby edge-on disk Galaxies. Such a synchrotron halo may serve as a tracer of 
the non-thermal particle population (CR population) at GeV energies. 
\textcolor{black}{This model corresponds to a traditional Leaky Box model, which is known to have problems when one tries to reproduce the observed composition
of CRs arriving at Earth, see for instance the discussion in \citet{StrongM2007}. In the future we will address this issue once include detailed calculations of CR spallation
and a Galactic wind.}
\textcolor{black}{In our model} CRs also escape if they cross the outer radius $r = r_{\rm max} = 20$ kpc of the Jansson-Farrar model.
For the more realistic parameters adopted below (the results shown in Figures 2 through 5) that happens infrequently.
\nskip
\textcolor{black}{It should be emphasized that all {\em determinations} of grammage are usually done by a careful analysis of the chemical composition of CR nuclei arriving 
{\em at Earth}, with the help of a nuclear network that describes spallation, decay etc. This is not what we do here: we calculate the accumulated grammage {\em at the moment of escape} from the Galaxy using a smooth matter distribution in the Galactic disk. The typical value for the grammage obtained in this fashion will
therefore tend to be slightly larger than the observed grammage. Nevertheless, the fact that CRs reside in the Galaxy for $\sim 10^8$ years and that -during that time- 
they diffuse inside the extended Galactic disk means that efficient mixing occurs in the direction perpendicular to the disk 
so that the grammage determined from CR nuclei arriving at Earth, 
the mean grammage of the whole CR  population and the grammage at the moment of escape are not too different. 
In future simulations we will address this point by explicitly calculating the grammage of the CR population in a local volume around the Solar system.}
\nskip
The residence time distribution for particles originating from a single source depends on the source location within the Galaxy, and in particular the Galacto-centric radius $r$. 
The residence time distribution for CRs originating from multiple sources at different \textcolor{black}{Galacto-centric} radii will be considered in a later paper. 
\nskip
To test our model calculation we show in Figure \ref{figure1} the  distribution of the CR particles injected in the mid-plane at a radius $r = 8\; {\rm kpc}$ on the positive $x$-axis.
Particle positions are projected onto the Galactic plane. 
The value for the ratio of the perpendicular and parallel diffusion coefficients was chosen equal to 
$\Dperp/\Dpar=10^{-6}$. This (unrealistically) small value for $\Dperp/\Dpar$ was chosen by purpose so that particle trajectories closely follow magnetic field lines.
In the simulations in this paper we choose a time step $\Delta t = 1,000 \; {\rm years}$, corresponding to a parallel diffusive step of order 
$\Delta s_{\parallel} = \sqrt{2 \Dpar \Delta t} \simeq 14 \; {\rm pc}$. 
We record the particle positions every 1000 time steps during the propagation, until they reach the upper or lower boundary at $ z=\pm H_{\rm cr}$ or $r = r_{\max} = 20$ kpc. 
The resulting distribution of CR positions is roughly equivalent to the case where the source has a constant strength in terms of the number of CR produced per unit time. 
One clearly sees the spiral structure of the Jansson-Farrar disk field in the particle positions since the
ratio of the typical diffusive step perpendicular to the field and along the field is small: $\Delta s_{\perp}/\Delta s_{\parallel} \simeq \sqrt{D_{\perp}/D_{\parallel}} = 10^{-3}$
for this simulation. The ring-like feature at $r = 20 \; {\rm kpc}$ is an artifact that results from our assumption that particles escape freely once they reach that radius and the fact that we record the final position of these escaping particles.
\nskip
To investigate the effects of the strength of perpendicular diffusion and of the injection radius on the residence time of CRs in the Galaxy we  have simulated the 
propagation for CR sources that are located at different radii in the Galactic plane, namely at $2\; \rm kpc$, $4\; \rm kpc$, $8\; \rm kpc$, and $10\; \rm kpc$. 
In each case we simulate four  different values for the ratio $\Dperp/\Dpar$:
we take $\Dperp/\Dpar \equiv \epsilon$ $ \;= 0.01,\; 0.1,\; 0.5,\; 1.0$. The last case corresponds to isotropic diffusion, 
where there is no influence of the direction of the magnetic field on the propagation of the particles.
The parallel diffusion coefficient is kept constant with $\Dpar = 3 \times 10^{28} \: {\rm cm}^2 \: {\rm s}^{-1}$, close to the typical value derived in models using
isotropic diffusion.
\nskip
In Figures \ref{figure2} though \ref{figure5} we show the spatial distributions of 
CR particles injected in the disk mid-plane at $2\; \rm  kpc$, $4\; \rm kpc$, $8\; \rm kpc$, and $10\; \rm kpc$, 
with $\epsilon=   0.01,\; 0.1,\; 0.5,\; 1.0$, respectively. The parallel diffusion coefficient is kept constant at $\Dpar = 3 \times 10^{28} \: {\rm cm}^2 \: {\rm s}^{-1}$
The four figures show that for the small value of $\epsilon$ the particles tend to follow the spiral disk field. For $\epsilon= 0.5$ 
the diffusion in the disk become more isotropic, while for $\epsilon = 1$ one gets pure isotropic diffusion where the magnetic field geometry no longer influences propagation.
\nskip 
In  Figure \ref{figure6} we show how the escaping CRs are distributed over the grammage when injected at a radius of $2\; \rm  kpc$, $4\; \rm kpc$, $8\; \rm kpc$, $10\; \rm kpc$, 
again for the four values of the diffusion ratio. 
The grammage obtained from the particles distribution  agrees with the grammage seen by typical CR. 
As one can see when the values of the diffusion ratio is small, the grammage distribution become larger, 
because the particles spend more time longer before they escape from the Galaxy. 
Figure \ref{figure7} shows  the residence time distribution for particles injected at a Galacto-centric radius  
$r = 2\; \rm  kpc$, $4\; \rm kpc$, $8\; \rm kpc$, $10\; \rm kpc$, and for the same four values of the ratio $\Dperp/\Dpar$. 
The residence time obtained from the distribution  agrees with the typical residence time  $t_{\rm res} \simeq \textcolor{black}{10^{8}}$ yr that is 
usually associated with CR propagation in the Galaxy in simulations like GALPROP and DRAGON.

\section{ Discussion \& Conclusions }
\label{sec:5}
In this work we have presented  a first test of a  simple  model that uses an approach based on the stochastic differential equations (in the It\^o formulation)  
to describe the CR propagation. We solve these equations numerically, using the simplest possible scheme \textcolor{black}{and the simplest possible model, the Leaky Box model where
CRs escape at the upper and lower edge of a CR halo.} This scheme takes account of the anisotropic diffusion 
that is expected in the presence of a large-scale, ordered 
Galactic magnetic field. For this Galactic field we have used the Jansson-Farrar model. The calculations presented here neglect the effects of the 
rigidity dependence of the diffusion coefficients by
neglecting energy losses while the CRs propagate.
\nskip
This method is  well-suited for simulating CR propagation in complex field geometries because it is fast and easy to implement 
numerically, compared with the more traditional approach that numerically solves the equivalent diffusion equation. The results presented here are for impulsive injection of the CRs and focuses on the effect of choosing different values for the ratio 
$\Dperp/\Dpar$, and the particle injection site. 
\nskip
Our results, and in particular Figure \ref{figure7}, show that the magnitude of $\epsilon = \Dperp/\Dpar$ has a major influence on the
residence time of the escaping CRs. Increasing $\Dperp$ (with fixed $\Dpar$) from $\Dperp = 0.01 \: \Dpar$ to $\Dperp = \Dpar$ (isotropic diffusion)  
significantly reduces both grammage and residence time by more than an order-of-magnitude. 
This behavior is a direct consequence of the field geometry in the Jansson-Farrar model and the
fact that we inject our CRs in the mid-plane of the Galactic disk, where the disk field is typically stronger than the
vertical component of the $X$-field. Most of the young stars that can act as the progenitors of Type II (core-collapse) supernovae, probable sources for Galactic CRs, 
reside in the thin stellar disk with a thickness $\sim 0.3$ kpc, (e.g. \citet{BinTre}). So as a first approximation injection at the mid-plane is reasonable.
Typically the ratio of the radial field component $B_{r}$ and the component $B_{\rm n}$ normal to the Galactic disk is $B_{\rm n}/B_{r} \simeq 0.1-0.3$, 
close to the mid plane at $r \sim 5$ kpc. 
This means that the effective diffusion coefficient in the direction perpendicular to the Galactic disk,
\ba
	D_{\rm n} & \equiv &  \hatn \bdot \bm{\mathrm{D}} \bdot \hatn = \Dperp \: \left[ 1 - (\hatb \bdot \hatn)^2 \right] + \Dpar \: (\hatb \bdot \hatn)^2
	\nonumber \\
	& & \\
	& = & \Dpar \: \left\{ \epsilon \: \left[ 1 - (\hatb \bdot \hatn)^2 \right] + (\hatb \bdot \hatn)^2 \right\} \; , \nonumber
\ea
with $\hatn$ the unit vector normal to the disk, 
is dominated by $\Dperp$ unless $\Dperp/\Dpar \ll (\hatb \bdot \hatn)^2/\left[ 1 - (\hatb \bdot \hatn)^2 \right] \ll 1$ for typical conditions.
\nskip
This also explains why for small $\Dperp$ CRs travel further along the spiral field lines, as is obvious from comparing the four different panels in 
Figures 2 through 5. 
In the simulations presented here the typical residence time in these cases is almost entirely determined by CR escape at the upper/lower edge of the cosmic ray halo so that 
the residence time is of order
\be
	t_{\rm res} = \frac{H_{\rm cr}^2}{2 D_{\rm n}} \; .
\ee
The typical distance traveled along the (curved) magnetic field in a residence time is
\be
	\Delta s_{\parallel}  \simeq \sqrt{2 D_{\parallel} \: t_{\rm res}} \simeq H_{\rm cr} \: \left(\frac{D_{\parallel}}{D_{\rm n}} \right)^{1/2} \; .
\ee
In fact, the models with $\Dperp/\Dpar = 0.5 \: , \: 1$, close to isotropic diffusion, produce
a significantly shorter residence time that the time $t_{\rm res} \simeq 10^8 \; {\rm years}$ indicated by observations for particles with GV rigidities. 
This is due to our choice of a constant $\Dpar \simeq 3 \times 10^{28} \; {\rm cm}^2 \: {\rm s}^{-1}$. By decreasing $\Dpar$ while keeping 
the ratio $\Dperp/\Dpar$ constant the residence time in these simple models would increase as $t_{\rm res} \propto \Dpar^{-1}$ without changing the
typical spatial distribution of the escaping CRs. In that sense our results are quite general for given $\epsilon$ and field geometry, 
and can therefore be scaled with the characteristic
time $t_{0} = H_{\rm cr}^2/2D_{\parallel}$.

\clearpage
\clearpage
\begin{figure}
\includegraphics[width=\columnwidth]{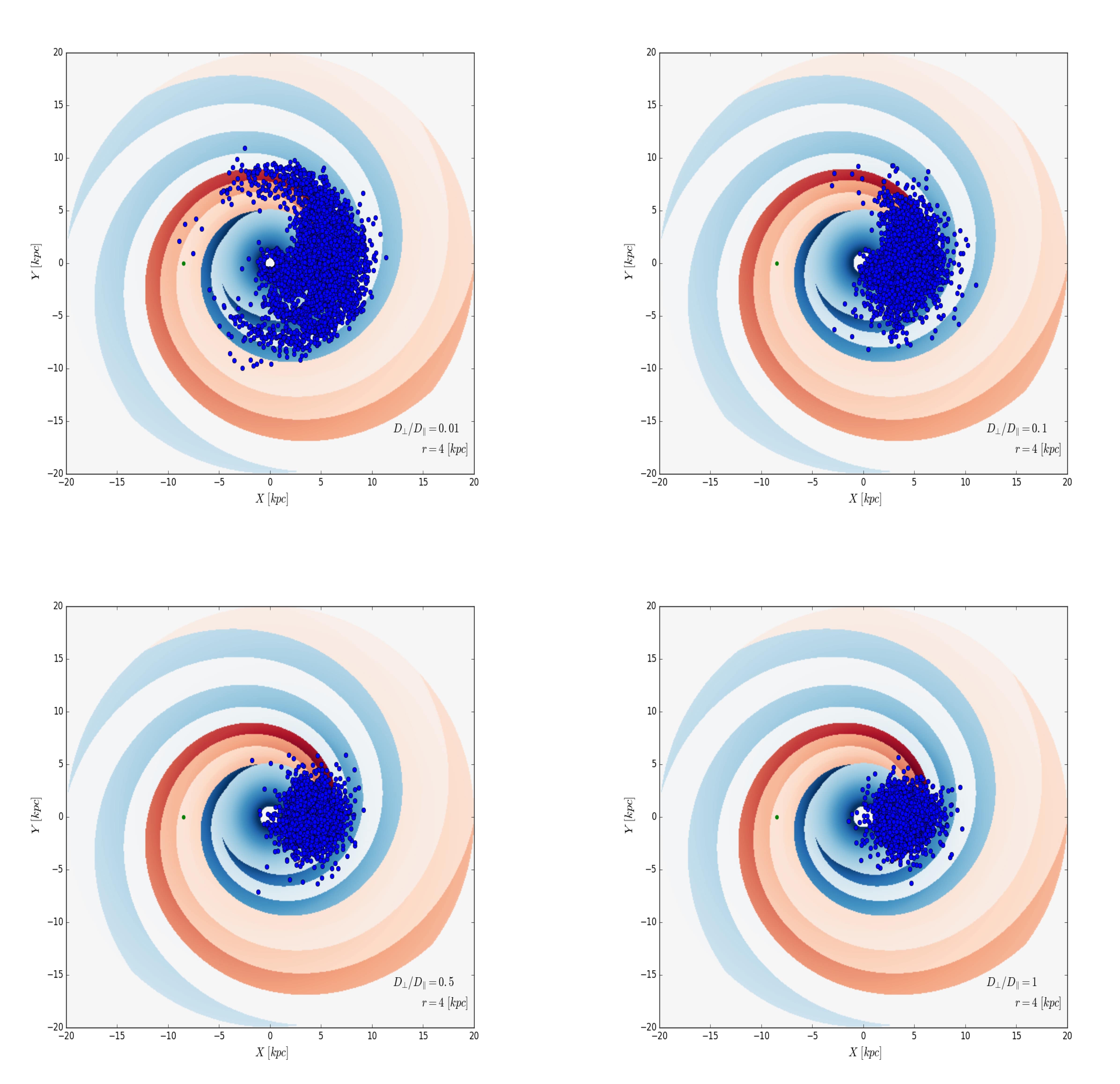}
\caption{As in Figure 2, for injection at a radius of 4 kpc.}%
\label{figure3}  
\end{figure}
\clearpage
\begin{figure}
\includegraphics[width=\columnwidth]{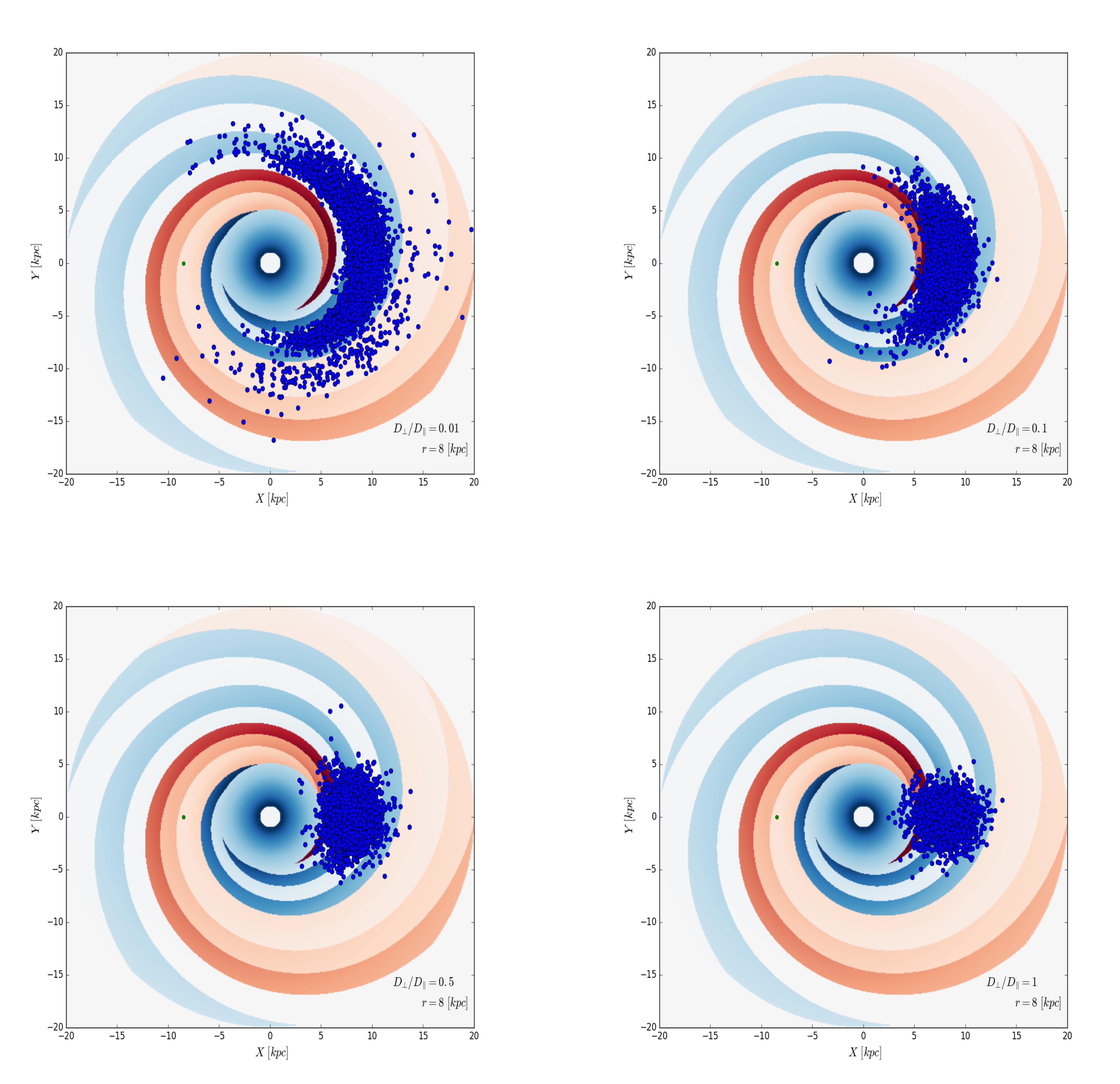}
\caption{As in Figure 2, for injection at a radius of 8 kpc. }%
\label{figure4}  
\end{figure}
\clearpage
\begin{figure}
\includegraphics[width=\columnwidth]{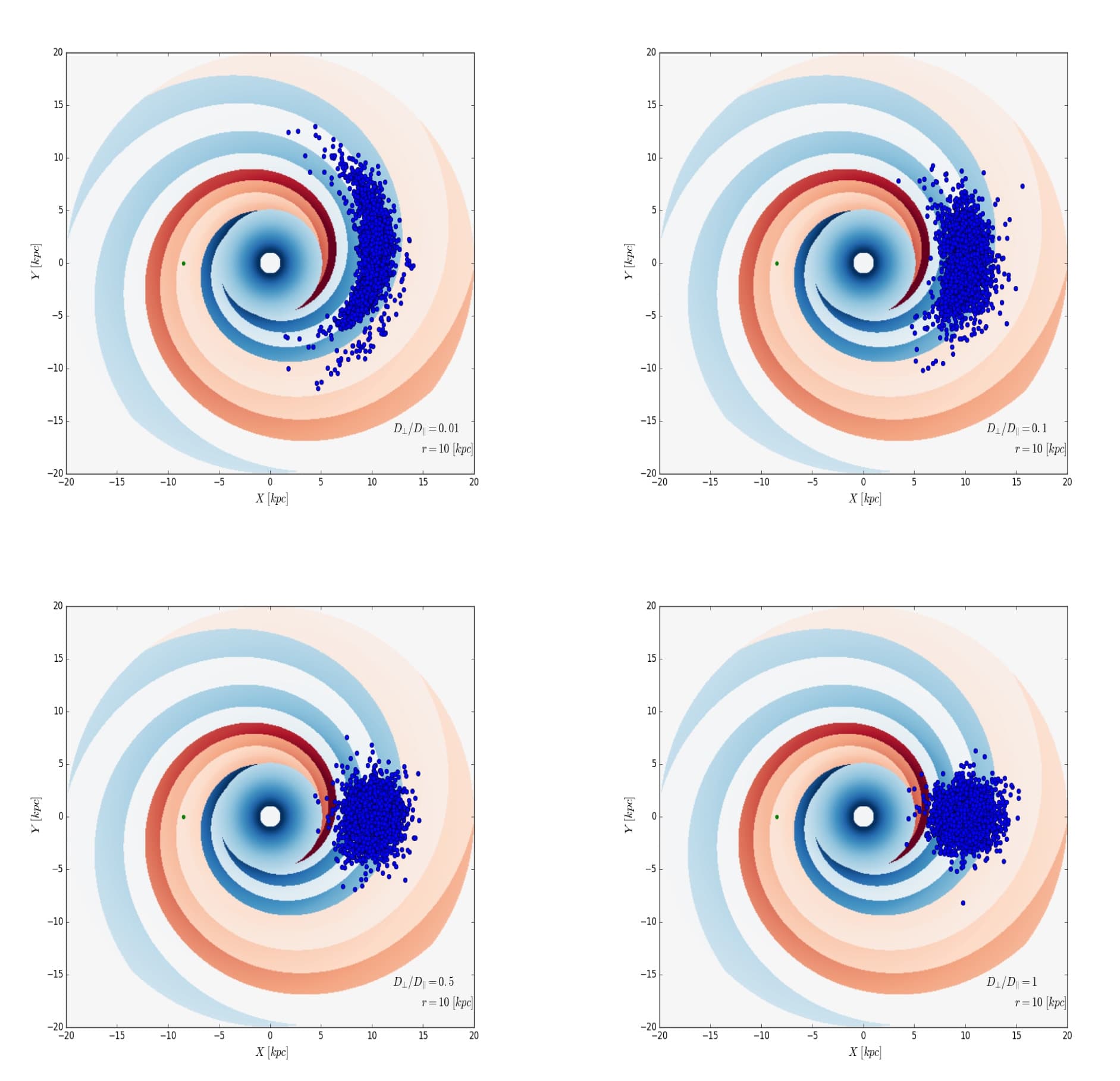}
\caption{As in Figure 2, for injection at a radius of 10 kpc. }%
\label{figure5}  
\end{figure}
\clearpage
\begin{figure}
\includegraphics[width=\columnwidth]{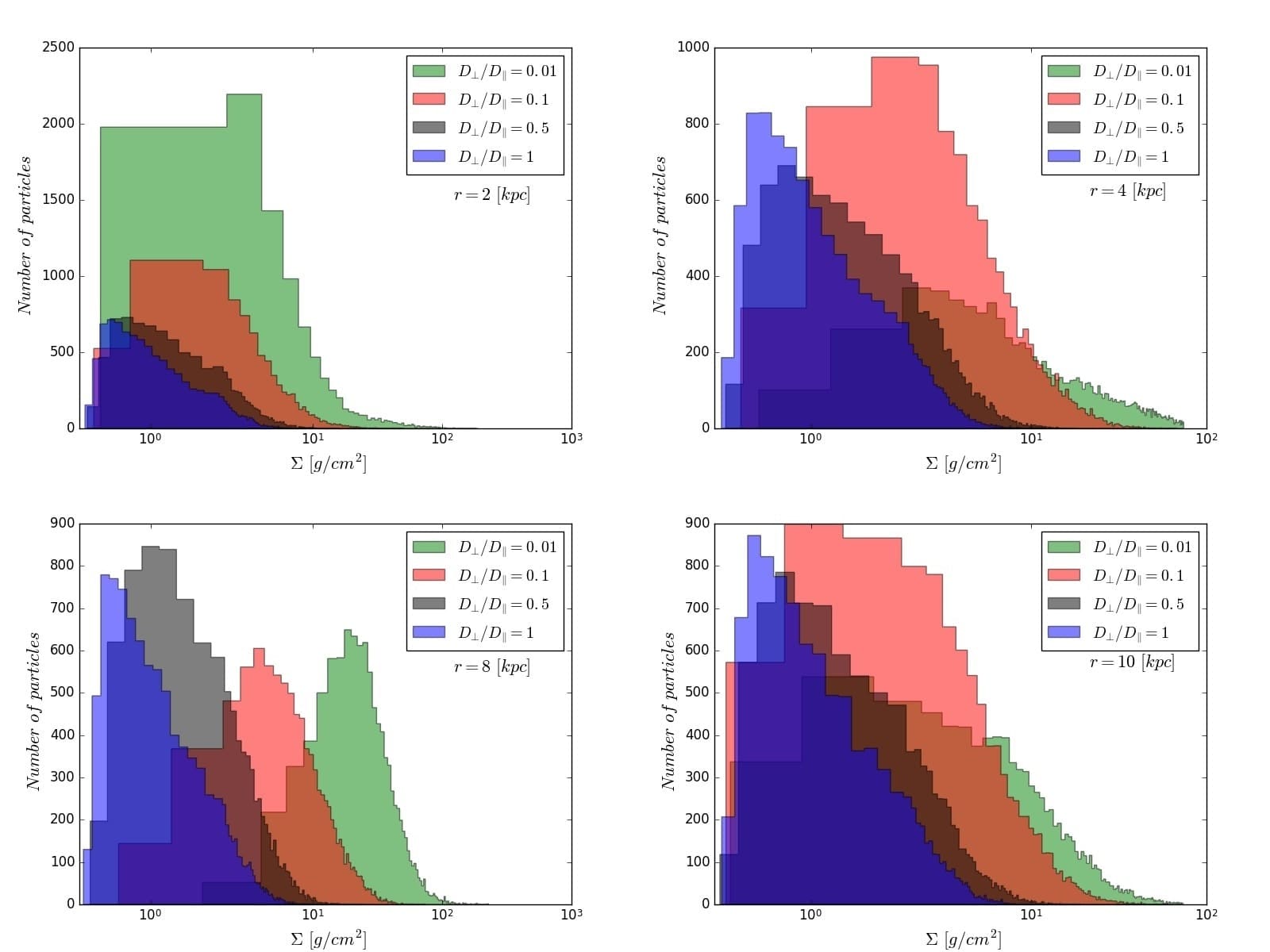}
\caption{The grammage distribution for the simulated cosmic rays at the moment of escape. 
The four panels are for an injection radius equal to $2\;\rm  kpc$, $4\;\rm kpc$, $8\; \rm kpc$ and $10\;\rm  kpc$ respectively. In each panel the
grammage distribution is shown for four values of the ratio $ \epsilon =\Dperp/\Dpar$: $\epsilon = 0.01 \: , \: 0.1 \: , \: 0.5$, and $1.0$,  
as indicated in each of the four panels. }%
\label{figure6}%
\end{figure}
\clearpage
\begin{figure}
\includegraphics[width=\columnwidth]{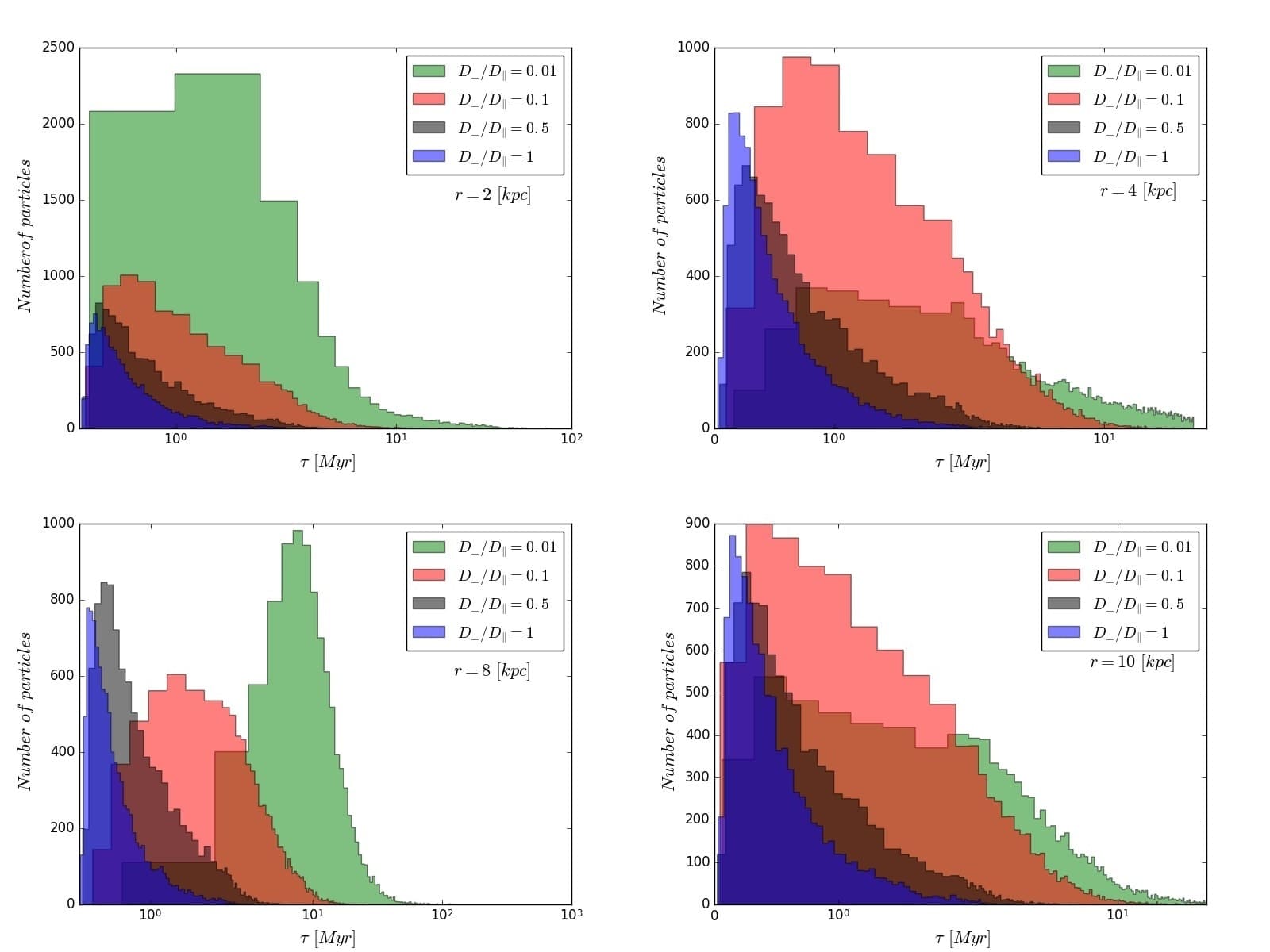}
\caption{ The residence time distribution for the simulated cosmic rays at the moment of escape. 
The four panels are for an injection radius equal to $2\;\rm  kpc$, $4\;\rm kpc$, $8\; \rm kpc$ and $10\;\rm  kpc$ respectively. In each panel the
residence time distribution is shown for four values of the ratio $ \epsilon =\Dperp/\Dpar$: $\epsilon = 0.01 \: , \: 0.1 \: , \: 0.5$, and $1.0$,  
as indicated in each of the four panels.}%
\label{figure7}%
\end{figure}

\clearpage
The effect of $\epsilon = \Dperp/\Dpar$ on the grammage distribution of the escaping CRs, shown in Figure 6, mostly reflects the variation
of the residence time as the typical radial scale length of the density distribution of the diffuse ISM (Eqn. \ref{cdens}) is $R_{\rm d} = 3.15$ kpc, and only varies
for $R > R_{\rm c} = 7$ kpc. Most particles in our simulation stay within $r = 10$ kpc, and see little (radial) variation of the diffuse gas density.  
We therefore conclude that the influence of anisotropic diffusion on the residence time is an important feature in determining the accumulated grammage of cosmic rays in the Galaxy.
\nskip
 In a forthcoming  paper we will use this model to study the residence time distribution for CRs originating from multiple sources, and  the 
interactions in the form of the spallation processes that occur during propagation when CR nuclei collide with atoms in the interstellar molecular and diffuse gas. 
We will also consider the effects of cross-field drift in the curved magnetic field and consider the effects of advection of CRs away from the disk by a Galactic wind.

\bibliographystyle{mnras}
\bibliography{Al_Zetoun_references}

\end{document}